# Implementation of transformer-based LLMs with large-scale optoelectronic neurons on a CMOS compatible platform

NEIL NA,[1,*] CHIH-HAO CHENG,[1] SHOU-CHEN HSU,[1] CHE-FU LIANG,[1] CHUNG-CHIH LIN,[1] NATHANIEL Y. NA,[1] ANDREW I. SHIEH,[1] ERIK CHEN,[1] HAISHENG RONG,[2] AND RICHARD A. SOREF[3]

[1]*Artilux Inc., Zhubei City, Hsinchu County 30288, Taiwan ROC*
[2]*Intel Corporation, Santa Clara, California 95054, USA*
[3]*Department of Engineering, The University of Massachusetts, Boston, Massachusetts 02125, USA*
**neil@artiluxtech.com*

**Abstract:** The recent rapid deployment of datacenter infrastructures for performing large language models (LLMs) and related artificial intelligence (AI) applications in the clouds is predicted to incur an exponentially growing energy consumption in the near-term future. In this paper, we propose and analyze the implementation of the transformer model, which is the cornerstone of the modern LLMs, with novel large-scale optoelectronic neurons (OENs) constructed over a complementary metal-oxide-semiconductor (CMOS) compatible platform. With all of the required optoelectronic devices and electronic circuits integrated in a chiplet only about 2 cm by 3 cm in size, 175 billon parameters in the case of GPT-3 are shown to perform inference at an unprecedented speed of 12.6 POPS using only 40 nm CMOS process node, orchestrated by an optoelectronic version of systolic array with no data skew and negligible propagation delay, along with a high power efficiency of 74 TOPS/W and a high area efficiency of 19 TOPS/mm$^2$. The influence of the quantization formats and the hardware induced errors are numerically investigated, and are shown to have a minimal impact. Our study presents a new yet practical path toward analog neural processing units (NPUs) to complement existing digital processing units.

## 1. Introduction

The recent boom of large language models (LLMs), mostly based on the transformer model in deep learning proposed by Vaswani et al. [1], has brought tremendous technological and social impacts. In particular, the astronomical number of training parameters needed, progressing from hundreds of billions [2] to more than one trillion [3] and beyond, boosts the construction of graphical processing units (GPU)-centric hyperscale data centers that daily consumes, e.g., in the case of Alphabet's Google, a whooping energy consumption of around 60-80 GWh in 2023 for inference only [4], and shows no sign of slowdown. The origin of such a huge energy consumption is at least two-fold: First, due to the growing energy cost about a few tens of pJ/bit of the server-to-server and rack-to-rack, electrical and optical, high-speed communication links [5] (this issue is beyond the scope of this paper); Second, due the stagnation of Moore's law in which the energy cost of multiply-and-accumulate (MAC), i.e., the main type of operation used in matrix multiplications essential for deep learning, has only been reduced to pJ/MAC level with digital gates implemented by complementary metal-oxide-semiconductor (CMOS) transistors [6]. Therefore, the exploration of new computing technologies to tackle the energy crisis is more urgent than ever.

In the literature, a variety of novel computing schemes based on processing photons have been proposed and studied, attracting wide attention from both academia and industry (see Ref. [7-10] for several review papers). These schemes can be generally categorized into, e.g., free-space optics (FSO) [11-13,15,20] vs. photonic-integrated circuit (PIC) [14,16-19], coherence-based [11,13-15,20] vs. intensity-based [12,16-19], involving additional degrees of freedom during a MAC operation, such as time [15,16], wavelength [18], radio frequency [19], angle and polarization [20], and a combination of all these techniques. While people have demonstrated encouraging results showing the feasibility of significantly improving the energy consumption of the MAC operation used in vector-vector, vector-matrix, and matrix-matrix multiplications, these implementations focus mostly on proof-of-concept small-scale image classification models. In fact, only a handful of papers analyze real-world large-scale language models [21-24] that are most pertinent to the energy crisis discussed in the previous paragraph.

In this paper, we propose and analyze a practical hardware implementation of the transformer-based LLM, using the demodulator (or lock-in) pixels originally designed for indirect time-of-flight (TOF) sensing and imaging applications [25,26], paired with appropriate emitter pixels, to function as optoelectronic neurons (OENs). The system is compatible with the commercial CMOS image sensor (CIS) hybrid bonding technology for three-dimensional sensing (3DS) [27-29], and the commercial chip-on-wafer-on-substrate (CoWoS) advanced packaging for GPU and high-bandwidth memory (HBM) [30]. Most importantly, with a chiplet size of only about 2 cm by 3 cm, an 8-bit LLM loaded with 175 billon parameters, i.e., in the case of GPT-3, can perform inference at the speed of 12.6 POPS, reaching a power efficiency of 74 TOPS/W and an area efficiency of 19 TOPS/mm$^2$ that both surpass those of the modern digital electronics [31] by roughly two orders of magnitude, all under realistic inclusion of the involved optoelectronic devices and analog frontend (AFE) electronic circuits, without resorting to bulky and costly implementations such as wavelength-division multiplexing (WDM). Note that our approach is different from the mainstream "tensor core" concept that breaks down the matrix multiplications into smaller units for a tensor core to calculate. Instead, our approach prioritizes the acceleration of giant matrix multiplications and massive parallelism, with smaller matrices compactly filling up a giant matrix multiplication for further calculations. Consequently, a substantial improvement of the power efficiency can be achieved for a large-scale neuron network that requires heavy parallel processing by pre-sharing the digital-to-analog converters (DACs), which will be shown to be the most power-hungry component in Sec. 4.1. Moreover, the demodulator pixels in the backside illumination (BSI) configuration, which will be discussed in details in Sec. 3.3, have been demonstrated to fit into a 10 μm-pitch 240×180 pixel array [27], a 5 μm-pitch 640×480 pixel array [28], and a 3.5 μm-pitch 1280×960 array [29], all vertically stacked to application specific integrated circuits (ASIC) through wafer-level hybrid bonding using 12″ Si wafers. Consequently, a substantial improvement of the area efficiency can be achieved, potentially outperforming conventional methods such as applying spatial light modulators (SLMs) in the case of FSO, and Mach-Zehnder interferometers (MZIs) or micro-ring resonators (MRRs) in the case of PIC.

The organization of this paper is as follows: In Sec. 2, the algorithm of the transformer-based LLM will be introduced and discussed, using GPT-3 with 175 billion weights as an example. In Sec. 3, the architecture of the OEN chip on the CMOS platform will be proposed and analyzed, including structural layouts and design considerations, temporal pipeline and timing, and modulation of photocurrent with demodulator pixel. In Sec. 4, the system evaluation and the emphasis on DAC will be examined, including key system performance metrics and their derivations, design and scaling of DAC through simulations, and summary of the calculated key performance metrics. In Sec. 5, the quantization formats considering integer number instead of floating-point number, and the hardware induced errors considering OEN variations, will be numerically experimented using visual transformer (ViT) models as examples. Finally, concluding remarks will be given in Sec. 6.

## 2. Transformer-based LLM – GPT-3 as an example

In this section, we take GPT-3 [3] as an example due to the relatively complete model information open to the public. Fig. 1 shows its structure, which loads a total number of weights of about 175 billion and is a transformer-based LLM. GPT-3 employs the self-attention mechanism inherent in the transformer model to extract the dependencies between input tokens. For example, at the last token inferencing step, 2048 tokens are first converted to model-recognized vectors by word embedding and position encoding. These tokens then pass through 96 layers of 96 multi-head self-attention (ATTN) modules and 1 feedforward (FF) network followed by layer normalizations. Finally, the predicted token is generated by word unembedding and Softmax. Massive amounts of vector-to-matrix multiplications (VMMs) are performed during the execution of the transformer model, and can be express as

$$y_i = a(\sum_j w_{ij} x_j + b_i) = a(\sum_k w_{ik} x_k) . \quad (1a)$$

$x_j$ is the input vector, $w_{ij}$ is the weight matrix, $b_i$ is the bias vector, and $y_i$ is the output vector. $a()$ is the activation function and can be linear or nonlinear depending on the usage. Here $b_i$ is absorbed by $w_{ij}$ to define $w_{ik}$, and the accumulations of each dot product $w_{ik}x_k$ can be regarded as a MAC. In the case where multiple $x_k$ are given, they can be concatenated along the column direction to form an input matrix for matrix-to-matrix multiplications (MMMs), and can be expressed as

$$y_{ij} = a(\sum_k w_{ik} x_{kj}) . \quad (1b)$$

Four main features of the transformer model can be observed: First, the MAC occupies the majority of the operations during the inference, corresponding to about 733 TO for the case of GPT-3 at the last token inferencing step. Second, massive parallelism in input vectors is adopted, and, compared to other deep learning algorithms such as convolution neuron networks (CNNs) and recurrent neuron networks (RNNs) that process one input vector at a time, the entire set of input vectors can be processed in parallel, corresponding to 2048 tokens for the case of GPT-3 at the last token inferencing step. In addition to the parallel processing in tokens, in the block of ATTN, the MMM with 96 sets of self-attention heads can also be processed in parallel. Moreover, each row of a weight matrix is capable of being processed in parallel: For example, at the last token inferencing step, the numbers of parallel tasks are 2048×(128×3×96) and 2048×12288 when constructing the query/key/value weight matrices $W_{K/Q/V}$ and the output weight matrix $W_{output}$ in the ATTN modules (128×3×96 and 12288 correspond to the row number of $W_{Q/K/V}$ times the set number of self-attention heads and the row number of $W_{output}$, respectively), and 2048×49152 and 2048×12288 when constructing the up-projection weight matrix $W_{up}$ and the down-projection weight matrix $W_{down}$ in the FF network (49152 and 12288 correspond to the row numbers of $W_{up}$ and $W_{down}$, respectively). Note from the perspective of hardware implementation, exploiting the possible parallel processing in a model can maximize the computing speed, but at the same time the computing power shall go up significantly unless the data are appropriately reused. This leads to the third feature of the transformer model, i.e., data reusing. Since the same-weight matrices in LLM such as $W_{K/Q/V}$, $W_{output}$, and $W_{up/down}$ are multiplied with different input vectors, the computing power can be suppressed by minimizing the times to read, convert, and apply the weight data from HBMs, by DACs, and to OENs, respectively. Therefore, how to densely integrate MAC computing units while leveraging the concepts of parallel processing and data reusing is the key to designing a powerful yet efficient hardware platform for LLM. Finally, while the values of $W_{K/Q/V}$, $W_{output}$, and $W_{up/down}$ are static, i.e., they are fixed after the training of the transformer model, the values of attention patterns are dynamic, i.e., they are updated whenever the next token is considered. Consequently, the MAC computing units must be capable of updating the weight values for matrix multiplications per clock cycle.

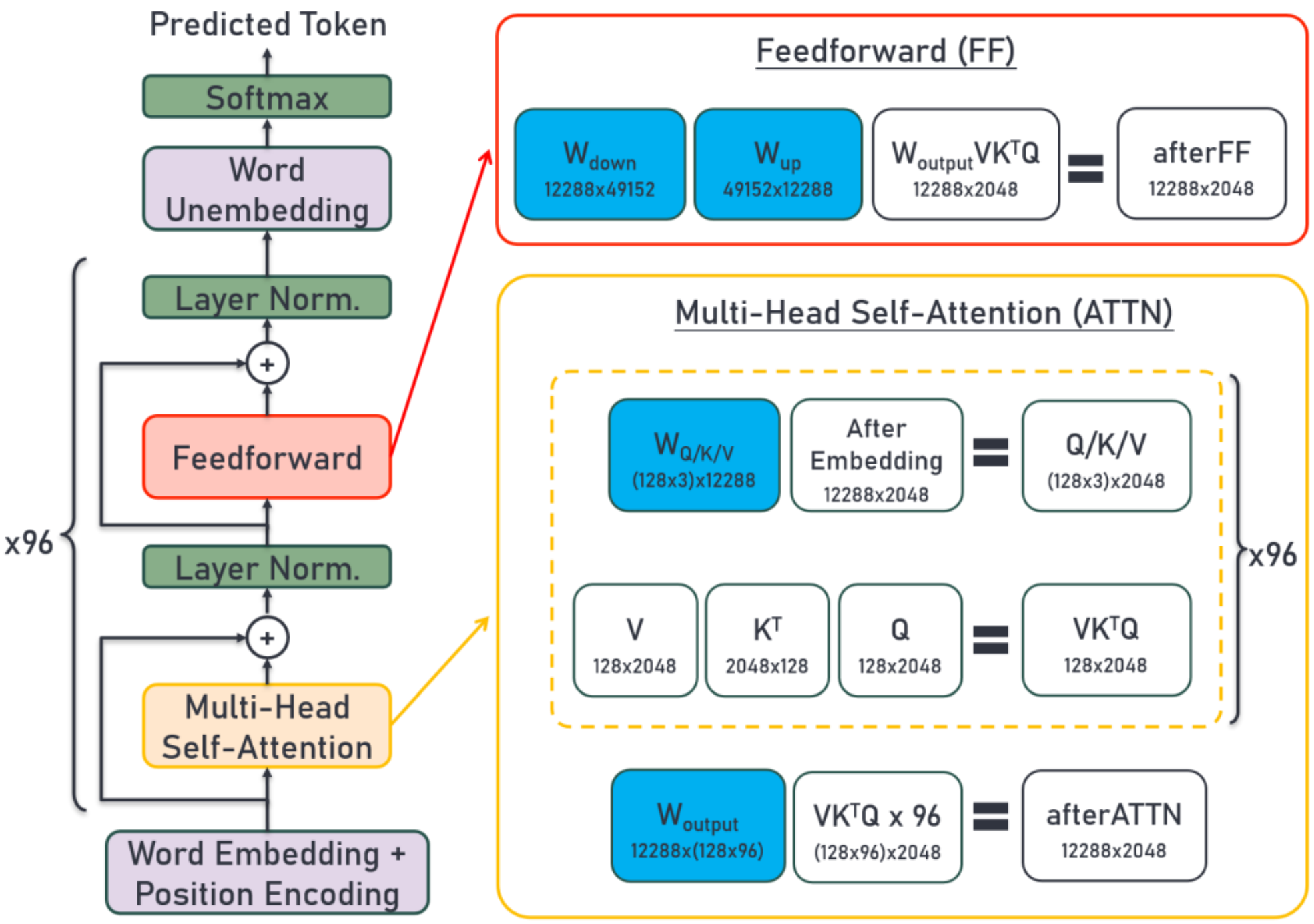

**Fig. 1.** The model structure of GPT-3, where the blue matrices are the pre-trained weights. The model mainly comprises 96 layers of 96 multi-head self-attention (ATTN) modules and 1 feedforward (FF) network. In the block of ATTN, the output after word embedding and position encoding is multiplied with 96 sets of self-attention heads containing weight matrices $W_{Q/K/V}$ to generate 96 sets of query/key/value Q/K/V matrices. In each self-attention head, the corresponding $K^TQ$ (self-attention pattern) is multiplied with V to form $VK^TQ$. 96 sets of $VK^TQ$ are then concatenated and multiplied with output weight matrix $W_{output}$ to generate the output of ATTN, i.e., afterATTN. In the block of FF, the output after multi-head self-attention and layer normalization is first up-projected by weight matrix $W_{up}$ and then down-projected by weight matrix $W_{down}$ (feature discrimination enhancement) to generate the output of FF, i.e., afterFF.

## 3. Architecture of the OEN chip on the CMOS platform

### *3.1 Structural layouts and design considerations*

Following the concept of the algorithms described in Sec. 2, we propose a highly-efficient neuron processing unit (NPU) chiplet as shown in Fig. 2, capable of processing MMM in a highly-parallel fashion. The chiplet consists of an OEN chip with three functional layers, several HBM chips placed at the sides of the OEN chip, and an Si interposer [30] that connects the OEN chip and the HBM chips. The three functional layers, including the illumination layer on the top, the sensing layer in the middle, and the processing layer at the bottom, are vertically stacked by wafer-level hybrid bonding. The structural layouts and design considerations are elaborated as follows.

#### 3.1.1 Illumination layer

The main function of the illumination layer is to send input vectors toward the sensing layer, where the value of each element in an input vector is encoded by modulating a different light intensity transmitted one by one at a time. By considering the token numbers of GPT-3, 2048 light sheets are needed to represent the 2048 tokens as the input vectors for maximizing the parallelism of the token dimension. Each light sheet should cover a group of demodulator pixels in the sensing layer where each of them is temporally modulated to serve as a weight vector. Since it is difficult to deploy lenses and diffractive optics to generate light sheets with a high aspect ratio (about 3000:1 needed) and a fine pitch (about a few microns needed), an emitter pixel array is adopted instead. Each emitter pixel in the illumination layer is one-by-one vertically aligned with each demodulator pixel in the sensing layer, and the emitter pixels located in the same row are connected in parallel by a routing wire and driven by the same DAC to form a pseudo light sheet.

As an example, benefitting from the recent advancement of fabrication and integration technologies, the GaN-based micro light-emitting diode (μLED) array [32] with GHz-range modulation bandwidths [33] can be fabricated at a small pitch ranging from a few to tens of microns, which makes the technology a good candidate for the illumination layer. The mature GaAs-based vertical-cavity surface-emitting laser (VCSEL) array could be another potential candidate.

#### 3.1.2 Sensing layer

The key element in the sensing layer is the demodulator pixel, which is originally designed for indirect TOF sensing and imaging applications where the phase shift and therefore the distance traverse of a modulated light beam can be detected by the demodulator pixels. For the computing applications discussed in this paper, due to the capability of operating beyond GHz speed, Ge-based demodulator pixels [25-28] are preferred over Si-based demodulator pixels [29] and deployed to serve the role of the MAC computing unit. Each of them detects the light intensities (i.e., receiving the input vector) from the emitter pixel, modulates the generated photocurrents one by one at each time step (i.e., serving the weight vector), and finally accumulates the total photocurrent-induced electron charges on the in-pixel capacitor during the whole illumination period (i.e., generating the dot product between the input vector and the weight vector). The detailed mechanism on how the demodulator pixel executes a MAC operation will be elaborated in Sec. 3.3. The row number of the (emitter pixel) array is chosen to be 2048 to be compatible with the number of pseudo light sheets. The column number of the (demodulator pixel) array is chosen to be 3072, because it is the common factor between the numbers of the possible parallel tasks when constructing the weight matrices of ATTN and FF. Such a choice also keeps the full chip within a reasonable size: By taking the 10 μm-pitch pixel array as an example, the full chip size is approximately equal to 3 cm by 2 cm, which is slightly smaller than the size of a standard full-frame sensor.

Leveraging from the data reusing feature in the transformer model, the demodulator pixels located in the same column are connected in parallel to share the same driving voltage from a single DAC, which saves the computing power substantially when the demodulator pixel loads are of sufficiently high impedance.

#### 3.1.3 Processing layer

The processing layer is filled with electronic circuits including the AFE, which bridge the optoelectronic devices in the illumination and sensing layers to the digital domain. To demonstrate the feasibility of adopting a mature and cost-effective process node on the CMOS platform for implementing the transformer-based LLMs, in this paper, the electronic circuits in this layer are designed and simulated at the clock rate of 1 GHz (unless stated otherwise) assuming 40 nm CMOS process node.

The center part of the processing layer is an analog-to-digital converter (ADC) array, which occupies an area that is the same size as the demodulator pixel array. Each ADC is dedicated to acquire the MAC results from a group of demodulator pixels directly sitting on top of it via Cu-Cu connection. Here we design each ADC having 8-bit data precision, 100 MHz sampling frequency, and 200 μm by 40 μm size, and then design an array of 154×512 ADCs where each one of them handles the readout of 20×4 demodulator pixels. Located in the north and the west part of the processing layer, there are the receiver (Rx) and the transmitter (Tx) one-dimensional (1D) DAC arrays that are connected to the demodulator pixels in the sensing layer and to the emitter pixels in the illumination layer by through-silicon vias (TSVs), respectively. Finally, the east part of the processing layer consists of digital circuits, such as 1) processors that perform minor operations such as activation, layer normalization, and Softmax, and 2) controllers and routers that not only synchronize and coordinate the operation timing of other circuits on the same layer, but also serve as the communication interfaces to HBMs and host devices.

#### 3.1.4 Memory and interconnect

There are two considerations on the memory required for the proposed OEN chip. First, the weights of GPT-3 should be loaded into on-chip memories, which account for 175 GB in an 8-bit format. Second, the data rate of on-chip memories should be high enough to support the readout rate of DACs (5.12 TB/s) and the write-in rate of ADCs (7.9 TB/s), such that there is no need of additional on-chip memory buffers to temporally store the data due to data rate mismatch. Unlike conventional GPUs that adopt complicated memory hierarchy, e.g., the mixture of L1/L2 cache, SDRAM, and HBM, to optimize the performance of memory access, here we adopt only one type of memory cell, i.e., HBM, to meet all the previously mentioned requirements while running of the OEN chip. HBM3e is the latest version of commercially available HBMs and each device can provide a capacity of 24 GB, a data rate of 1.2 TB/s, and an access energy of 3.4 pJ/bit [34-36]. Therefore, 8 HBM3e chips should be integrated on the Si interposer to work in conjunction with the OEN chip. Moreover, before directly loading the weights to the 8 HBMs chips, the weights originally trained with floating-point format (FP) should be converted to 8-bit integer format (INT8) with a post-training quantization (PTQ) algorithm or a quantization-aware training (QAT) algorithm, and then properly distributed to the 8 HBM3e chips to maximize the overall memory bandwidth.

Regarding the interconnects between the host devices and the NPU chiplet, there are two cases that should be considered. Conventionally, in the beginning of each inference, the tokens in the form of text format (UTF or ASCII) are converted to the word embedding vectors by lookup tables (LUTs). When the conversion is processed by the LUTs stored in the NPU chiplet, since the total data size before word embedding is only about a few kB, any standard interconnect should be sufficient to complete the token transmission in a negligible time. When the conversion is processed by the LUTs stored in the host devices, since the total data size after word embedding is approximately 25 MB, the data transfer will lead to a latency about 400 μs via PCIe 5 (32 GT/s per lane; maximal 16 lines), which is still negligible compared to the entire inference time to be calculated in Sec. 4.3.

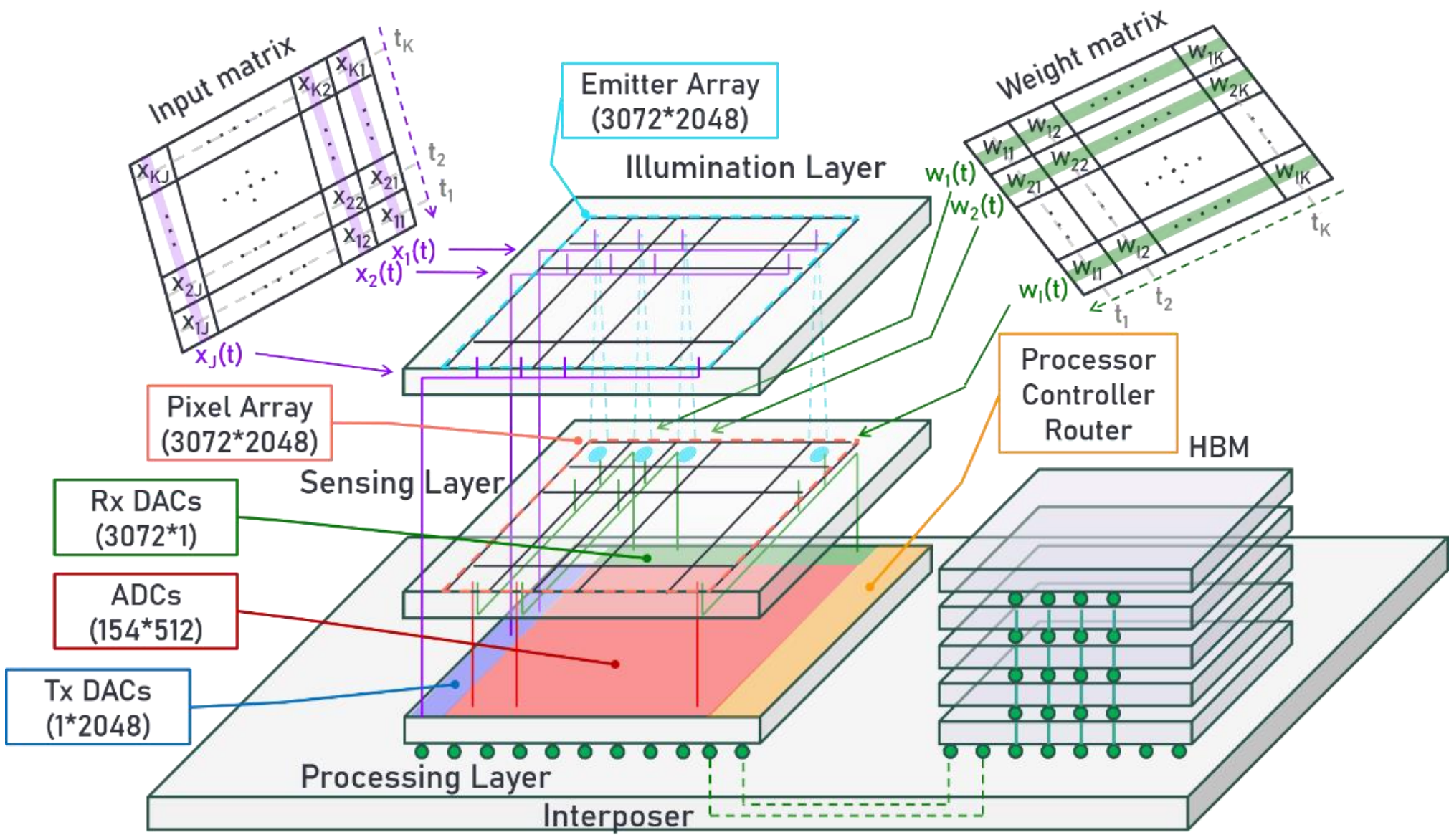

**Fig. 2.** The schematic plot of the neuron processing unit (NPU) chiplet, where a single optoelectronic neuron (OEN) chip is connected to multiple high-bandwidth memories (HBMs) via a Si interposer. The OEN chip can be decomposed into the illumination, sensing, and processing layers, which are vertically stacked by wafer-level hybrid bonding. See Sec. 3.1 for detailed descriptions on the emitter pixel array, demodulator pixel array, transmitter/receiver (Tx/Rx) digital-to-analog converter (DAC), and analog-to-digital converter (ADC). The processing of the matrix-to-matrix multiplication (MMM) expressed in Eq. (1b) is illustrated at the same time.

### *3.2 Temporal pipeline and timing*

Fig. 3 shows the temporal pipeline of the NPU chiplet, in which the operation flow of each component and their temporal interactions are illustrated. At the start of an inference, the tokens from the host devices are converted to the word embedding vectors and are temporally stored on HBMs as a preparation for the following utilization. Once the preparation is completed, one temporal repeat of a MMM computation is initiated. First, the processors, together with controllers, send the read commands to HBMs to stream out input vectors and weight vectors, and drive them out synchronously via Tx DACs and Rx DACs to execute point-wise multiplications of MMM. During this illumination-and-sensing period, the results of point-wise multiplications of MMM are also accumulated on the in-pixel capacitors, and last for a time equal to the vector length multiplied by the inverse of the clock frequency to drive the DACs. Taking the length of the word embedding vectors 12288 as an example and assuming 1 GHz clock rate, the illumination-and-sensing period is about 12 μs. Second, the accumulated MMM results are held and sampled by ADCs. Assuming all ADCs operate at 100 MHz sampling frequency, and each one of them handles the accumulated MMM results from 20×4 demodulator pixels in sequence, the readout time is equal to 800 ns. Meanwhile, the processors and the controllers would send the write commands to HBMs to save these held and sampled MMM results to a proper memory space on HBMs. After the readout process is completed, each in-pixel capacitor is reset to a reference voltage and then waits for the next temporal repeat of the MMM computation. Finally, depending on the model structure, the stored MMM results are optionally read from HBMs to the processors to compute operations other than MAC, such as layer normalization and Softmax, and then written back to HBMs from the processors for further utilization.

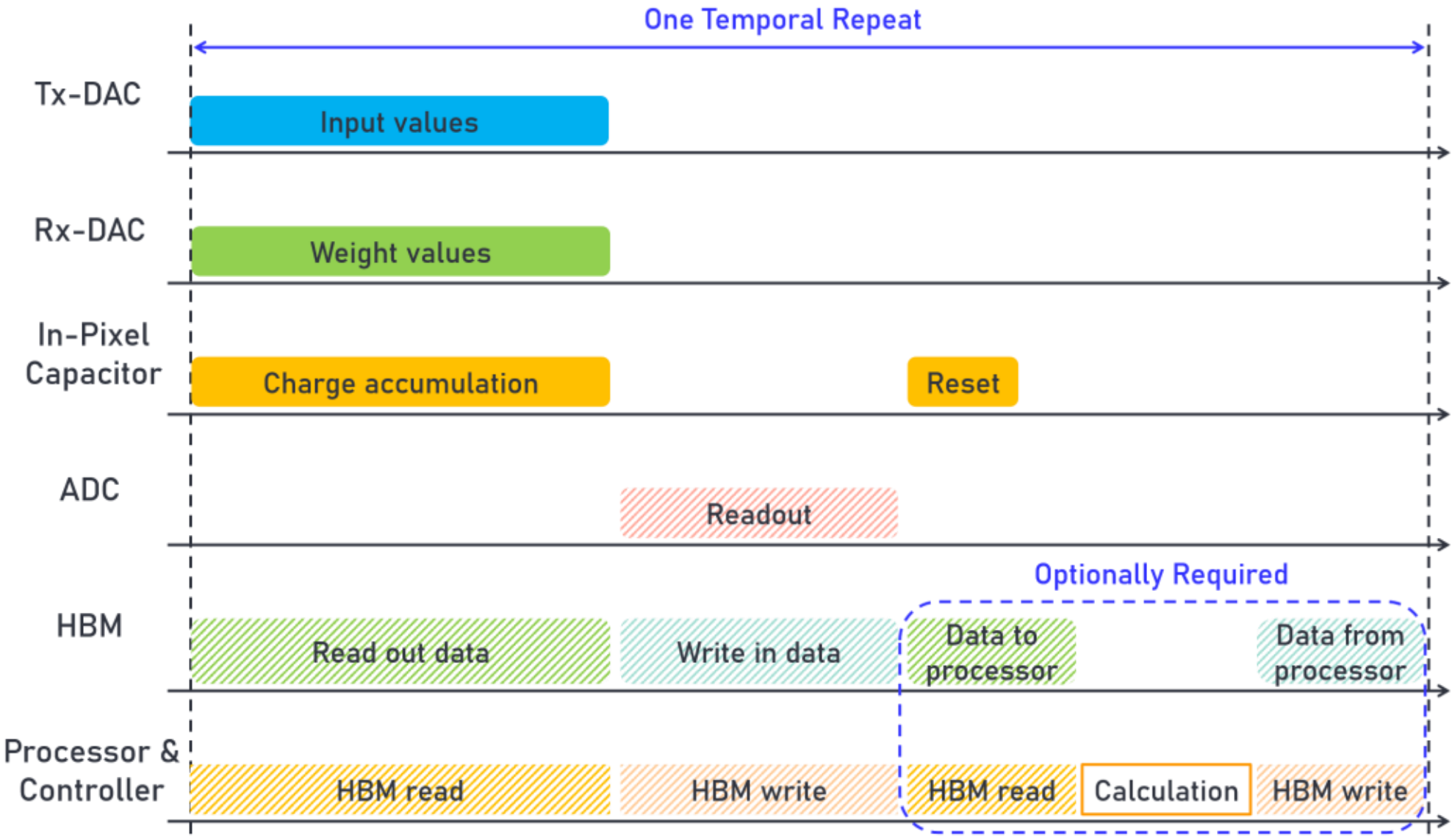

**Fig. 3.** The timing diagram of the proposed neural processing unit (NPU) chiplet. The flow and the interaction of its key components including transmitter/receiver (Tx/Rx) digital-to-analog converters (DACs), in-pixel capacitors, analog-to-digital converters (ADCs), and high-bandwidth memories (HBMs), are shown to perform one temporal repeat of a matrix-to-matrix multiplication (MMM) computation.

### *3.3 Modulation of photocurrent with demodulator pixel*

Fig. 4 shows the schematic plot of a two-tap demodulator pixel (see the inset in the box) and the timing diagram on how to perform a MAC operation by applying a bipolar complementary processing sequence to it. The two-tap demodulator pixel can be treated as one photodiode connected to two (or multiple) switches followed by two (or multiple) in-pixel capacitors. However, instead of using two transistors connecting to one photodiode, the device is commonly constructed using the resistor-based structure or the capacitor-based structure, in which the device surface terminals are implemented by the "N+, P+, P+, N+" contacts (the P+ terminals are for the resistor-based gate control; the N+ terminals are for collecting the photo-electrons generated between the two N+ terminals) or the "N+, MOS, MOS, N+" contacts (the MOS terminals are for the capacitor-based gate control; the N+ terminals are again for collecting the photo-electrons generated between the two N+ terminals) [29], respectively. During the bipolar complementary processing sequence, two sub-cycles are prepared. The value (and its complementary value) of the photo-response from the one effective photodiode in the first sub-cycle (and in the second sub-cycle), and the values (and their complementary values) of the gate controls from the two effective switches in the first sub-cycle (and in the second sub-cycle), are multiplied and accumulated, leading to the tap+ and tap− demodulation signals arisen from the collected electron charges on the in-pixel capacitors. The resultant voltages from the source-follower transistors (not shown in Fig. 4) after the in-pixel capacitors are fed to the ADCs as the desired differential-mode voltage input.

The purpose of the bipolar complementary processing sequence is two-fold: First, through the bipolar complementary processing sequence, the original unipolar input and weight signals are converted to bipolar input and weight signals to facilitate the capability of manipulating both positive and negative numbers in MMMs. Second, while ultra-low dark current has been achieved in the assumed the GeSi materials [37,38], the device inevitably generates some dark current and, without the complementary processing sequence, the resultant voltages are fed to the ADCs as the unwanted common-mode voltage input. Note that by adding two more sub-cycles, it is possible to construct a symmetrized bipolar complementary processing sequence to additionally eliminate the asymmetry between the tap+ and tap− demodulation signals due to fabrication and integration imprecisions.

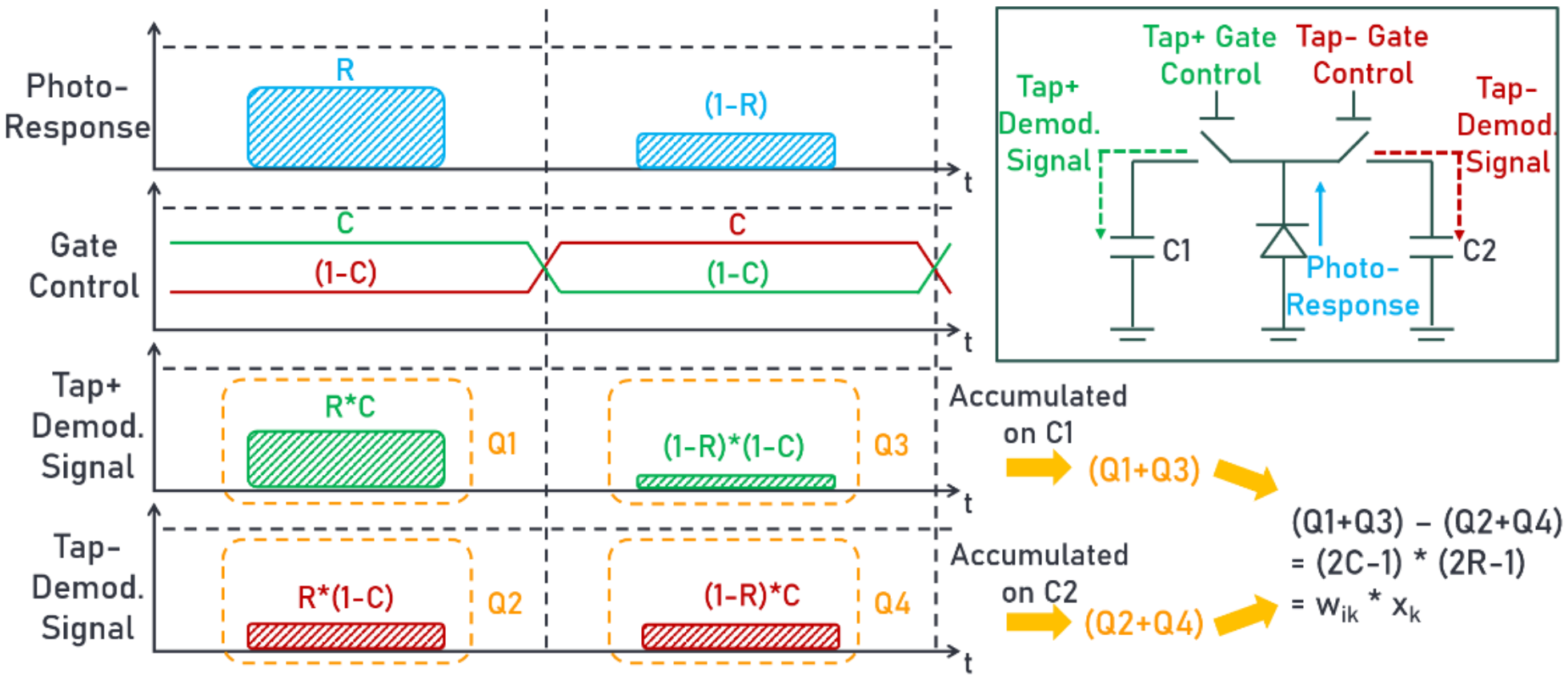

**Fig. 4.** The timing diagram of the complementary processing sequence applied to the two-tap demodulator pixel. The schematic plot of the two-tap demodulator pixel is shown in the inset in the box. Note that values of photo-response R and gate control C ∈ [0,1] so that values of input (2R-1) and weight (2C-1) ∈ [-1,1].

## 4. System evaluation and the emphasis on DAC

### 4.1 Key system performance metrics and their derivations

In this section, the performance of running the entire GPT-3 model using the proposed NPU are comprehensively analyzed. The key model parameters are first defined as follows: $T$ and $L$ are the numbers of tokens (2048) and layers (96) for the whole model, respectively. In the block of ATTN, $H$ is the number of heads (96), and $S$ and $N$ are the dimensions of key/query/value (128) and output (12288), respectively. In the block of FF, $M$ and $N$ are the dimensions of neuron (49152) and word embedding (12288), respectively. Then, the key hardware parameters are defined as follows: $f_{clk}$ is the clock frequency. $C_T$ and $C_W$ are the numbers of row and column of the pixel array, respectively. When the size of pixel array cannot handle all input and/or weight vectors at a time, i.e., when $C_T$ is smaller than $T$ and/or $C_W$ is smaller than the number of parallel workloads for weight vectors, some numbers of temporal repeats defined by $R_T$ and/or $R_W$ are required to complete the total tasks for inputs and weights, respectively. $r$ represents the number of sub-cycles that is described in Sec. 3.3, e.g., 1, 2 and 4 for the processing sequence that is unipolar single, bipolar complementary, and symmetrized bipolar complementary, respectively.

#### 4.1.1 Computing speed

To analyze the computing speed, the total number of the operations for the inference with the maximum number of tokens in GPT-3 is evaluated. According to the model structure introduced in Sec. 2, here we only consider the MAC operations that process the MMMs in ATTN and FF because they account for most of the computations about 733 TO. Note that since the formation of the self-attention pattern $K^TQ$ using key vectors K and query vectors Q as well as its interaction with value vectors V through $VK^TQ$ only amount to about 20 TO, they can be safely neglected to simplify the equations to be derived in the following. By including only the number of MACs that process the MMMs between the input tokens and the weight matrices $W_{Q/K/V}$, $W_{output}$, and $W_{up/down}$, the overall system operation tasks can be written as

$$\begin{aligned} n_{sys} &= 2\left(\left(3\cdot SH\times N+N\times SH\right)+\left(M\times N+N\times M\right)\right)\cdot TL \\ &\underset{SH=N}{=} 2\left(4N^2+2MN\right)TL \end{aligned}, \quad (2)$$

where the factor of 2 is used convert the number of MACs to the number of operations. By considering the number of parallel workloads in MMMs using the $C_T$×$C_W$ pixel array, the overall system operation delay can be written as

$$\begin{aligned} \tau_{sys} &= \left(f_{clk}^{-1}Nr\frac{3\cdot SH}{C_W}+f_{clk}^{-1}SHr\frac{N}{C_W}+f_{clk}^{-1}Nr\frac{M}{C_W}+f_{clk}^{-1}Mr\frac{N}{C_W}\right)\cdot R_T L \\ &\underset{SH=N \text{ and } R_T=\frac{T}{C_T}}{=} f_{clk}^{-1}\left(4N^2+2MN\right)r\frac{TL}{C_T C_W} \end{aligned}. \quad (3)$$

Note that since the readout time of ADCs, estimated to be 800 ns as mentioned in Sec. 3.2, is much shorter than the exposure time, i.e., $Nr/f_{clk}$, $SHr/f_{clk}$, and $Mr/f_{clk}$, the readout time of ADCs is not included in $\tau_{sys}$. By further dividing the overall system operation tasks $n_{sys}$ in Eq. (2) by the overall system operation delay $\tau_{sys}$ in Eq. (3), the computing speed can be derived as

$$\gamma = 2 f_{clk} r^{-1} C_T C_W . \quad (4)$$

4.1.2 Computing power efficiency and system power

To analyze the power related performance, we start with breaking down the energy consumption of the proposed NPU. To complete the VMM/MMM as described in Eq. (1a)/(1b), the energy consumption can be divided into three parts, i.e., input energy $E_x$, weight energy $E_w$, and output energy $E_y$. These three energy parts can be further decomposed into

$$E_x = E_{EM} + E_{DAC|EM} + E_{read} \quad (5)$$

$$E_w = E_{DM} + E_{DAC|DM} + E_{read} \quad (6)$$

$$E_y = E_{ADC} + E_{write} , \quad (7)$$

where $E_{EM}$, $E_{DM}$, and $E_{ADC}$ are the energy consumptions of an emitter pixel, a demodulator pixel, and an ADC, respectively; $E_{DAC|EM}$ and $E_{DAC|DM}$ are the energy consumptions of a DAC to drive an emitter pixel and a demodulator pixel, respectively; $E_{read}$ and $E_{write}$ are the energy consumptions for a memory read and write, respectively. In Eq. (5), owing to the small impedance of the emitter, it is safe to assume $E_{DAC|EM}$ efficiently transfers to $E_{EM}$ with little energy loss, and then a portion of $E_{EM}$ transfers to the light pulse energy according to the emitter power conversion efficiency. Moreover, there is a minimal light pulse energy required to overcome the photo and dark current shot noises in the demodulator pixel. Therefore, Eq. (5) can be re-written as

$$E_x = E_u(T_{expo}) + E_{read} , \quad (8)$$

where $E_u$ is the minimal unit pulse energy needed to drive an emitter pixel while having a sufficient signal-to-noise ratio (SNR) at the same time, and is a function of the exposure time $T_{expo}$ to complete a series of operations. To derive the lower bound of $E_u$, we consider the condition that the quantization error of the analog signal should be at least equal or greater the analog noise [39]. Here the analog signal and analog noise refer to the mean value and standard deviation value of the differential charges accumulated on the in-pixel capacitors, respectively, and such a condition can be expressed as

$$\frac{\frac{1}{q}\left(I_{max}-I_{min}=2I_{avg}\right)\alpha_{EM}\times T_{expo}}{\sqrt{12}\left(2^{b}-1\right)}\geq\sqrt{\frac{1}{q}\left(I_{avg}\alpha_{EM}+I_{dark}\right)\times T_{expo}}\ . \tag{9}$$

$q$ is the unit electron charge; $I_{max}$, $I_{min}$, and $I_{avg}$ are the maximum, minimum, and average photocurrents, respectively; $\alpha_{EM}$ is the duty cycle of the emitter; $b$ is the bit number of the ADC; $I_{dark}$ is the dark current. Note the factor √12 is due to the assumption of a uniform quantization error probability density function. By relating $I_{avg}$ and $E_u$ with

$$I_{avg}\alpha_{EM}=\frac{q}{\hbar\omega}\eta_{DM}\eta_{EM}\frac{E_u\times f_{clk}T_{expo}}{T_{expo}}, \tag{10}$$

substituting Eq. (10) into Eq. (9), and letting $T_{expo}=f_{clk}^{-1}Nr$, (9) can be re-written as

$$E_u\geq\varepsilon\cdot\delta\ \text{ with }\ \varepsilon=\begin{cases}\dfrac{3\hbar\omega}{\eta_{DM}\eta_{EM}}(2^{b}-1)^{2}\\[2ex]\dfrac{3\hbar\omega}{\eta_{DM}\eta_{EM}}\sqrt{\dfrac{I_{dark}}{3qf_{clk}}}(2^{b}-1)\end{cases}\ \text{ and }\ \delta=\begin{cases}\dfrac{1}{Nr}\\[2ex]\dfrac{1}{\sqrt{Nr}}\end{cases}, \tag{11}$$

for the limiting cases when

$$\begin{cases}I_{dark}<<I_{th}\\ I_{dark}>>I_{th}\end{cases}\ \text{ with }\ I_{th}=\frac{3q}{4f_{clk}^{-1}Nr}\left(2^{b}-1\right)^{2}. \tag{12}$$

$\eta_{DM}$ and $\eta_{EM}$ are the demodulator quantum efficiency and the emitter power conversion efficiency, respectively; $I_{th}$ is the threshold dark current for determining the limiting cases. As expected, due to the Poissonian nature of the analog signal and analog noise, a longer $T_{expo}$ results in a lower $E_u$. Note that given 2 GHz clock rate, $r$=2, and $b$=8, $I_{th}$ ranges from 78 nA to 0.78 nA given $N$ between $10^2$ and $10^4$, and is much larger than the typical dark currents reported for both state-of-the-art Ge-based and Si-based demodulator pixels.

Based on the above energy definitions, and considering the total energy consumption within the overall system operation delay in the ATTN modules, the system power of the block of ATTN can be written as

$$\begin{aligned}P_{sys}&=\frac{\left(E_x\left(3\cdot SH\times Nr+N\times SHr\right)+E_w\left(3\cdot SH\times Nr+N\times SHr\right)+E_y\left(3\cdot SH+N\right)\right)TL}{\tau_{sys}}\\&\underset{SH=N}{=}\left(\left(\varepsilon\cdot\delta+\frac{E_{read}}{C_W}\right)+\left(E_{DM}+\frac{E_{DAC|DM}+E_{read}}{C_T}\right)+\frac{E_{ADC}+E_{write}}{Nr}\right)f_{clk}C_TC_W\\&\underset{N>>1}{\approx}\left(\frac{E_{read}}{C_W}+\left(E_{DM}+\frac{E_{DAC|DM}+E_{read}}{C_T}\right)\right)f_{clk}C_TC_W\end{aligned}, \tag{13}$$

Note that in deriving Eq. (13), the energy consumptions in Eq. (6)-(8) and the minimal unit pulse energy in Eq. (11) are applied. Following the same approach, the system power of the block of FF can be written down by considering the total energy consumption within the overall system operation delay in the FF network, which results in the same equation as Eq. (13). Therefore, Eq. (13) also represents the system power for the entire GPT-3 model. By further dividing the computing speed $\gamma$ in Eq. (4) by the system power $P_{sys}$ in Eq. (13), the computing power efficiency can be derived as

$$\eta_p=\frac{2r^{-1}}{\dfrac{E_{read}}{C_W}+\left(E_{DM}+\dfrac{E_{DAC|DM}+E_{read}}{C_T}\right)}. \tag{14}$$

Now it can be observed that the energy consumptions of the memory read, the demodulator pixel, and the DAC to drive the demodulator pixel, are the main three factors determining the computing power efficiency. Moreover, benefitting from the feature of pre-sharing the DACs in the proposed NPU, the computing power efficiency is enhanced as the row and column numbers of the array scale up to larger values. Note that $E_{DM}$ is relatively small compared to $E_{DAC|DM}$ but it becomes the limiting constraint of the computing power efficiency for an infinitely large pixel array. In practice, it is difficult to design the DACs to keep $E_{DAC|DM}$ independent of the number of pixels being shared. Therefore, the energy consumptions of $E_{DM}+E_{DAC|DM}/C_T$ as a function of shared pixel number will be evaluated through detailed circuit simulations in Sec. 4.2 to accurately calculate the computing power efficiency.

#### 4.1.3 Computing area efficiency and system area

Considering the NPU chiplet as described in Sec. 3.1, the system area can be written as

$$A_{sys} = A_{pixel} C_T C_W + A_{DAC}\left(C_T + C_W\right) + A_{other}\,, \tag{15}$$

where $A_{pixel}$, $A_{DAC}$, and $A_{other}$ are the areas of the demodulator pixel, the DAC, and other circuits, e.g., the processor/controller/router in the OEN chip and the memory in HBM chip. By further dividing the computing speed $\gamma$ in Eq. (4) by the system area $A_{sys}$ in Eq. (15), the computing area efficiency can be derived as

$$\eta_a = \frac{2 f_{clk}}{\dfrac{A_{pixel}}{r^{-1}} + \dfrac{A_{DAC}}{r^{-1}\dfrac{C_T C_W}{C_T + C_W}} + \dfrac{A_{other}}{r^{-1} C_T C_W}}\,. \tag{16}$$

As expected, since the demodulator pixels occupy the area to handle massive amounts of MACs, $A_{pixel}$ is the main factor to determine the computing area efficiency when the row and column numbers of the array scale up to larger values. Another related parameter is the system power handling, which is defined as the system power $P_{sys}$ in Eq. (13) divided by the system area $A_{sys}$ in Eq. (15), a ratio that serves as an indicator that if cooling should be introduced to the NPU chiplet.

### *4.2 Design and scaling of DAC through simulations*

Here we evaluate two different DAC architectures: The first one, i.e., the R-2R DAC also known as RDAC, is shown in Fig. 5(a). It uses a series-parallel arrangement of controlled resistors to generate a specific voltage. To ensure the voltage stability, i.e., to ensure that the voltage output does not change with changes in the external impedance, the internal impedance must be significantly smaller than the external impedance for a good design. Therefore, when the external loading is light (heavy), a relatively larger (smaller) resistance should be selected, which consumes much less (more) energy. This means that every time the load doubles, one needs to reduce the internal resistance by half to maintain the voltage stability. Such a constraint can be observed in Fig. 5: When evaluating the total energy consumption, it increases linearly with the number of shared pixels as shown in Fig. 5(c), but the per-pixel energy consumption remains unchanged as shown in Fig. 5(d). Since the total energy consumption of RDAC becomes very large for a large pixel array, an alternative approach should be considered.

The second DAC architecture, i.e., the current-steering DAC also known as IDAC, is shown in Fig. 5(b). This architecture generates a specific voltage by switching multiple constant current sources. Although IDAC performs worse than RDAC in terms of energy consumption under light external loading, which is mainly due to the requirement of additional biasing circuits to control the outputs of the constant current sources, the number of additional biasing circuits increase slowly with the loading. Such a property can be observed in Fig. 5: When evaluating the total energy consumption, it increases slightly with the number of shared pixels as shown in Fig. 5(c), but the per-pixel energy consumption decreases substantially with the number of shared pixels as shown in Fig. 5(d).

Furthermore, there are area considerations as shown in Fig. 5(e). Because RDACs achieve a higher current demand by connecting resistors in parallel, their area increases linearly with the current demand. On the other hand, IDACs, while meeting mismatch requirements, achieve a higher current demand by adjusting the MOS aspect ratio, allowing their area to remain constant with the current demand. Finally, since giant matrix multiplications and massive parallelism are inherent in the transformer-based LLM, we choose the IDAC architecture to calculate the key system performance metrics in the next section, in order to minimize the computing power efficiency and maximize the computing area efficiency at the same time.

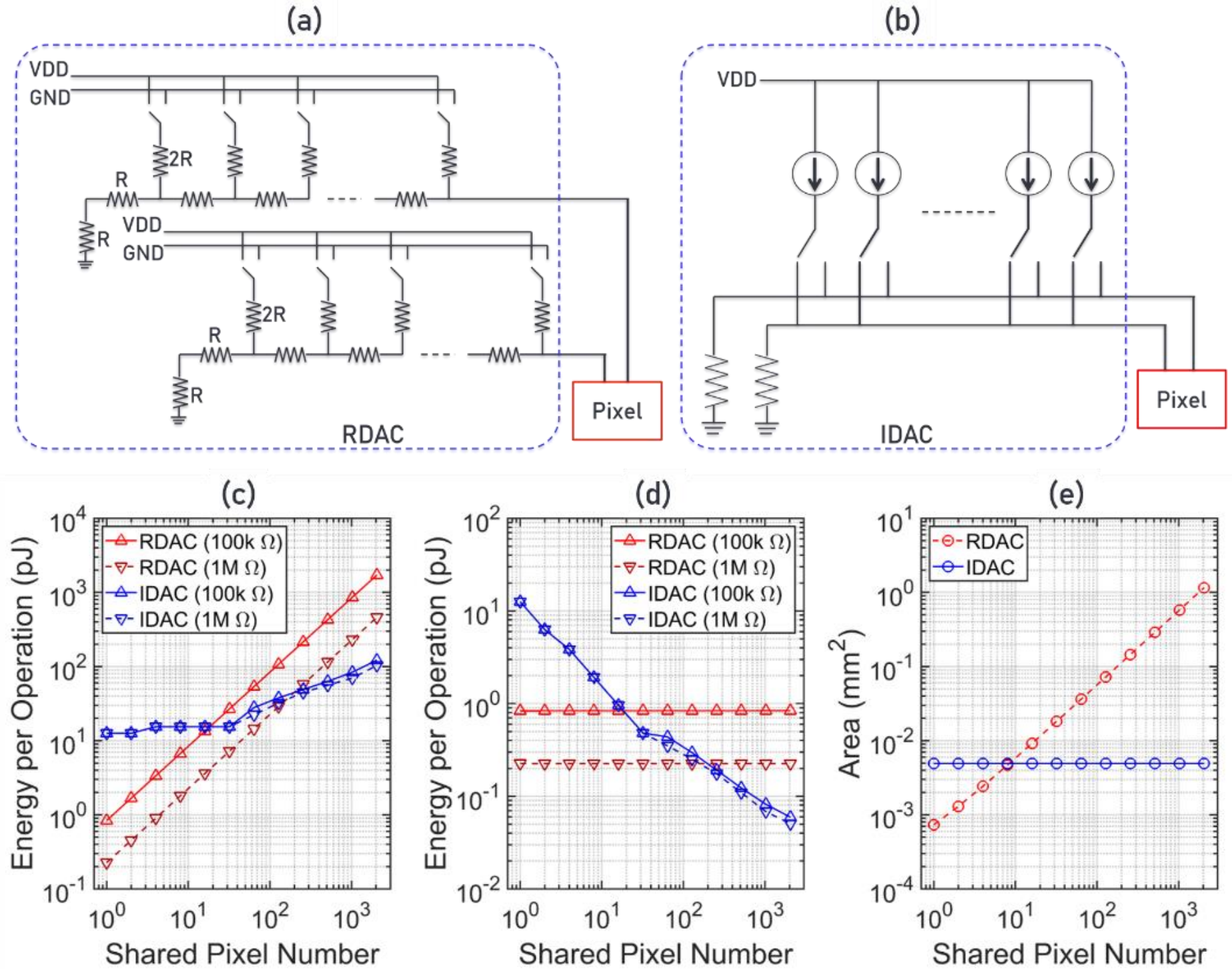

**Fig. 5.** The schematic plots of (a) R-2R digital-to-analog converter (DAC), i.e., RDAC and (b) current-steering DAC, i.e., IDAC, where VDD and GND are the supply voltage and ground, respectively. (c) The total energy consumption and (d) the per-pixel energy consumption, of 8-bit DACs, to drive different numbers of shared pixels in parallel. The red and blue colors correspond to RDAC and IDAC, respectively; the upward-pointing and the downward-pointing triangles correspond to the demodulator pixel loads of 100k Ω and 1M Ω, respectively. (e) The areas of RDACs (red circles/dashed line) and IDACs (blue circles/solid line) to drive different numbers of shared pixels in parallel. The simulations are done assuming 1 GHz clock rate.

### *4.3 Summary of the calculated key system performance metrics*

To calculate the performance of the proposed NPU to run an entire GPT-3 model, the simulation results of the IDAC obtained in Sec. 4.2 are taken into the formula derived in Sec. 4.1 for a complete analysis. The performance color maps in Fig. 6 illustrate the overall system operation tasks, computing speed, computing power efficiency, computing area efficiency, overall system operation delay, system power, system area, and system power handling, calculated as a function of the size of the pixel array, where $f_{clk}$ and $r$ are assumed to be 2 GHz and 2, respectively, and the demodulator pixels are assumed to be 10 μm in pitch. Note that to have a fair comparison with the publicly accessible specifications of GPUs, the consumed powers and the chip areas of the assumed HBM3e chips integrated on the proposed NPU are excluded from the analysis. As expected, the speed, power, and area, increase when the row or column number of the array scale up. More importantly, the power efficiency and the area efficiency increase significantly when scaling up the array in terms of the row number and the row or column number, respectively. Specifically, the key performance metrics of the proposed NPU for the implementation of 2048 rows and 3072 columns of pixels are listed in Table 1, which achieves a speed of 12.6 POPS, a power efficiency of 74 TOPS/W, and an area efficiency of 19 TOPS/mm$^2$, corresponding to a power of 172 W, an area of 654 mm$^2$, and a power handling of 262 mW/mm$^2$. Note the IDAC energy consumption here is further reduced by a factor of 1.85 due to the optimized IDAC design specific for the implementation of 2048 rows. Compared to the specifications of a single unit of Nvidia T4 [31], the proposed NPU shows superior performance in computing speed, computing power efficiency, and computing area efficiency by roughly two orders of magnitude. Although a hundred units of Nvidia T4 can be clustered on a rack to reach a similar computing speed as the proposed NPU, it is inevitable to pay the price of higher system power and larger system area by also roughly two orders of magnitude.

The reasons behind the high power and area efficiencies of the proposed NPU should be emphasized: First, the high computing power efficiency stems from the distinctive feature of large-scale energy sharing in the spatial domain, which can be accomplished when the IDAC architecture is adopted. Second, the high computing area efficiency stems from the unique property to compute the MAC operations in the time domain using the micron-size demodulator pixel on the CMOS platform, which circumvents the implementation of large-area adder trees as required in conventional digital electronics. Currently, LLMs are mostly executed by GPUs through cloud computing due to the huge power and area requirements. However, with the NPU performance predicted in this paper, edge computing for both training and inference becomes possible and is crucial to the commoditization of AI technologies.

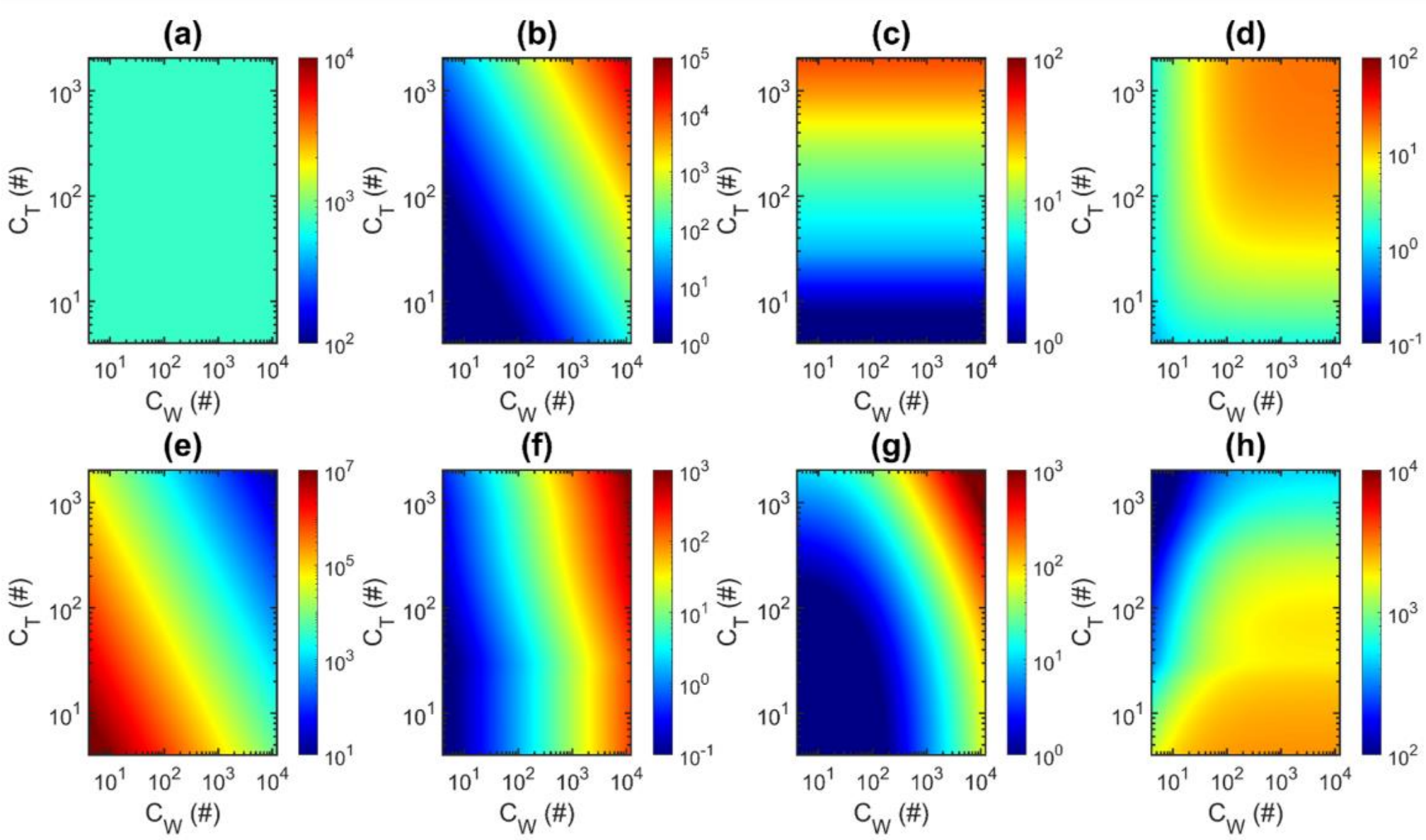

**Fig. 6.** The (a) overall system operation tasks (TO), (b) computing speed (TOPS), (c) computing power efficiency (TOPS/W), (d) computing area efficiency (TOPS/mm$^2$), (e) overall system operation delay (ms), (f) system power (W), (g) system area (mm$^2$), and (h) system power handing (mW/mm$^2$) of the proposed neural-processing unit (NPU)

running the entire GPT-3 model as a function of array format, where $C_T$ and $C_W$ are the numbers of pixels in the row and column directions, respectively.

**Table 1. Comparison of key system performance metrics.**

| INT8 | Tasks (TO) | Speed (TOPS) | Power Efficiency (TOPS/W) | Area Efficiency (TOPS/mm$^2$) | Delay (ms) | Power (W) | Area (mm$^2$) | Power Handling (mW/mm$^2$) |
|---|---|---|---|---|---|---|---|---|
| **OEN[a]**<br>**$C_T$=2048**<br>**$C_W$=3072** | 712 | 12583 | 74 | 19 | 57 | 172 | 654 | 262 |
| **Nvidia T4[b]**<br>**× 100** | 712 | 13000 | 0.32 | 0.24 | 54.77 | 40625 | 54166 | 750 |
| **Nvidia T4[b]**<br>**× 1** | 712 | 130 | 0.32 | 0.24 | 5477 | 406.25 | 541.66 | 750 |

[a] simulated using 40 nm CMOS process node.

[b] fabricated using 12 nm CMOS process node.

## 5. Quantization formats and hardware induced errors

To assess the impact of hardware-induced errors due to the variation of OENs during the execution of the transformer-based LLMs, we employ lighter ViT models [40, 41] of different number of parameters, i.e., ViT-Base, Small, and Tiny, without the loss of generality. ViT models, constructed upon the transformer architecture and utilized for image classification, serve as an appropriate benchmark to assess the impact of hardware-induced errors on the classification accuracy. The mini-ImageNet1K dataset is utilized, comprising 100 distinct classes with 120 images per class allocated for training and 50 images per class reserved for testing. Such a dataset, while being compact, retains sufficient diversity to probe the robustness of the ViT models across all classification categories. By integrating the framework of OEN into the computational pipeline of the ViT models, we quantify how errors affect the ViT models by characterizing the classification accuracy, providing critical insights into the viability of implementing transformer-based LLMs with OENs.

For the proposed and analyzed OEN chip on the CMOS platform, the hardware-induced errors are mainly attributed to quantization loss and device non-uniformity. Quantization loss originates from the finite precision of DAC and ADC, and, to address this effect, the algorithm LLM.INT8 [42] is adopted in the case of PTQ to emulate the impact of quantization on the matrix multiplications in the ViT models, including those in the ATTN modules and in the FF network. In this algorithm, activations, weights, and outputs are quantized from FP16 to INT8 precision, reflecting the resolution of the converter hardware. Values exceeding a predefined threshold (i.e., set to be 6 as default) are treated as outliers and are processed with higher precision FP to mitigate the information loss, while the remaining inlier values are quantized from FP16 to INT8 precision. Device non-uniformity, on the other hand, is introduced to reflect the spatial variations of the performance of the photonic and electronic components over the array. This effect is modeled through a Gaussian-distributed multiplicative noise, independently applied to both weights and activations. The standard deviation of the Gaussian distribution physically represents, e.g., the variations of the power conversion efficiency of the μLEDs/VCSELs, the variation of the quantum efficiency and the demodulation contrast of the demodulator pixels, and other possible factors in the OENs. These noises are incorporated into all matrix multiplications, thereby emulating the cumulative impact of the variations of OENs on the classification accuracy.

When PTQ is applied directly with INT8 precision, the evaluation results are illustrated in Fig. 7(a), where the noise strength or the standard deviation $\sigma$ of OENs is varied from 0% to 10%. The ViT-Base model exhibits superior resilience to noise, in which the classification accuracy degrades only by 1.5% (from 81.7% to 80.2%). In contrast, under the same conditions, the ViT-Tiny model exhibits inferior resilience to noise, in which the classification accuracy degrades as large as 8% (from 71% to 63%). This disparity suggests that larger models with more parameters, such as ViT-Base model, possess enhanced robustness to hardware-induced errors, implying that an increased number of OENs contributes to greater stability. Subsequently, QAT is applied through fine tuning with INT8 precision, to further mitigate the information loss due to quantization. After the original ViT-Tiny model is fine-tuned over the mini-ImageNet1K dataset, as illustrated in Fig. 7(b), the new ViT-Tiny model achieves an improved classification accuracy reaching up to 87% and 86.5% when $\sigma$ equals to 0 and 10%, respectively, showing a small degradation less than 2%.

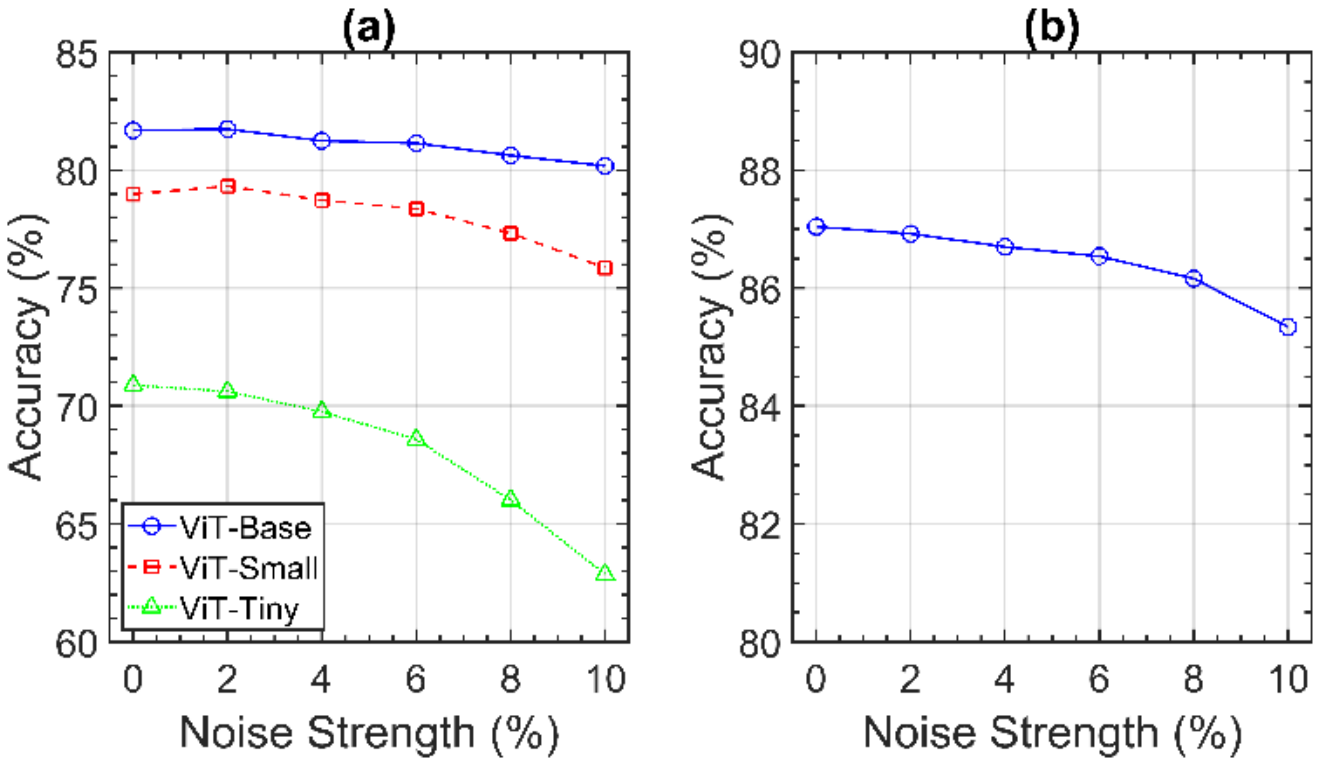

**Fig. 7.** (a) Classification accuracies of the visual transformer (ViT) models with INT8 post-training quantization (PTQ), plotted as a function of the optoelectronic neuron (OEN) noise strength. Results for ViT-Base (blue circles/solid line), ViT-Small (red squares/dashed line), and ViT-Tiny (green triangles/dotted line) models are shown. (b) Classification accuracies of the ViT-Tiny model with INT8 quantization-aware training (QAT), plotted as a function of the OEN noise strength.

## 6. Concluding remarks

When evaluating the key system performance metrics in Sec. 4.3, we carefully avoid the common issues in the literature: First, the clock rate under discussion is assumed to be around 1-2 GHz, which is typical for computation applications, instead of unrealistic clock rates such as 10-50 GHz or more by referencing high-speed optical fiber communication systems, which is in fact the other major efficiency bottleneck in datacenters due to the high baud rates. Second, regardless of the type of novel optical, photonic, or optoelectronic analog computing schemes, electrical circuits and interfaces are still indispensable to design and control the systems, and so any report excluding the contributions of electrical components such as DAC, ADC, etc., cannot claim accuracy in evaluating the key system performance metrics.

Some final comments on the proposed NPU: First, the MAC operations are physically performed through the manipulation of electrons rather than photons, i.e., by modulating the injection current in the emitter pixel and the photocurrent in the demodulator pixel. These choices are intentional, in order to maximize the power and area efficiencies with technologies that can be practically scaled up, because 1) using weak non-resonant interactions with photons lead to inefficient and large devices such as SLMs and MZIs, and 2) using strong resonant interactions with photons lead to devices requiring extreme fabrication precision and/or active feedback control such as MRRs. Second, one may ask why not further remove the optoelectronic emitter-demodulator configuration and replace it with a pure electronic source-switch configuration using CMOS transistors? The answer lies in Eq. (11), showing $E_u$ as low as a few photon energies can be used for each MAC operation. Instead, if CMOS transistors are used in implementing the (current) source and switch, the corresponding $E_u$ would be roughly four orders of magnitude higher because the on current is usually at the order of 1 μA (subthreshold currents suffer from variations exponentially relating to the threshold voltage mismatch and may not be suitable for analog computing). Moreover, implied in Eq. (11), the additional flicker and thermal noises associated with CMOS transistors can be treated as an effective $I_{dark}$ and further spoil $E_u$. Third, the giant matrix multiplication approach in this paper can only be achieved by a spatial array with time-based degree of freedom introduced, such as our NPU shown in Fig. 2 (spatial) and Fig. 3 (temporal), otherwise the instantaneous power can easily overwhelm any standard power supply due to without spreading the computation energy consumption over the time domain. In this sense, the proposed NPU may be treated as an optoelectronic version of systolic array with no data skew needed and negligible propagation delay across all pixels per clock cycle (i.e., the worst-case RC delay is around 100-150 ps, which is much smaller than the inverse of the clock frequency, considering differential backend metal traces typically of 50 Ω and 0.1 fF/μm). Finally, the improved power and area efficiencies when scaling up the array, as shown in Fig. 6, are in stark contrast to the tensor core approach, where both efficiencies would remain the same or even worse for the additional data communications between the multiple tensor cores.

**Acknowledgments.** N. Na would like to thank Mr. Y.-J. Lin, Dr. C.-Y. Chen, Dr. Y.-C. Lu, and Mr. T. Shia for their literature surveys.

**Disclosures.** All authors affiliated with Artilux are shareholders of Artilux Inc.

**Data availability.** The data that support the findings of this study are available within the article.